\documentclass[structabstract]{aa}
\usepackage[varg]{txfonts}
\usepackage{graphicx}
\usepackage{natbib}

\begin{document}

\title{Dynamical mass of the O-type supergiant in Zeta Orionis A\thanks{Based 
  in part on observations collected at the European
  Southern Observatory, Chile 
	(Prop. No. 076.C-0431, 080.A-9021, 083.D-0589, 285.D-5042)
  }}

\author{
  C.A.~Hummel\inst{1} \and 
  Th.~Rivinius\inst{2} \and
  M.-F. Nieva\inst{3,4} \and 
  O.~Stahl\inst{5} \and
  G.~van~Belle\inst{6} \and
  R.T. Zavala\inst{7}
}

\institute{
  European Southern Observatory,
  Karl-Schwarzschild-Str.~2, 85748 Garching
  \thanks{Correspondence: {\tt chummel@eso.org}}
  \and
  European Southern Observatory,
  Casilla 19001, Santiago 19, Chile 
  \and
  Dr.~Karl Remeis-Sternwarte \& ECAP, University of Erlangen-Nuremberg,
  Sternwartstr.~7, 96049 Bamberg
  \and
  Institute of Astro- and Particle Physics, University of Innsbruck,
  Technikerstr.~25, 6020 Innsbruck, Austria
  \and
  ZAH, Landessternwarte Heidelberg-K\"onigstuhl, 69117 Heidelberg
  \and
  Lowell Observatory, 1400 W. Mars Hill Rd., Flagstaff, AZ 86001, USA
  \and
  U.S. Naval Observatory, Flagstaff Station, 10391 W. Naval Obs. Rd., 
  Flagstaff, AZ 86001, USA
}

\date{Received: $<$date$>$; accepted: $<$date$>$; \LaTeX ed: \today}  

\abstract
{}
{A close companion of $\zeta$ Orionis A was found in 2000 with the Navy
Precision Optical Interferometer (NPOI), and shown to be a physical
companion.  Because the primary is a supergiant of type O, for which
dynamical mass measurements are very rare, the companion was observed with
NPOI over the full 7-year orbit. Our aim was to determine the dynamical
mass of a supergiant that, due to the physical separation of more than
10 AU between the components, cannot have undergone mass exchange with the 
companion.}
{The interferometric observations allow measuring the relative positions
of the binary components and their relative brightness. The data collected
over the full orbital period allows all seven orbital elements to be
determined.  In addition to the interferometric observations, we have
analyzed archival spectra obtained at the Calar Alto, Haute Provence,
Cerro Armazones, and La Silla observatories, as well as new spectra
obtained at the VLT on Cerro Paranal.  In the high-resolution spectra
we identified a few lines that can be associated exclusively to one or
the other component for the measurement of the radial velocities of
both. The combination of astrometry and spectroscopy then yields the
stellar masses and the distance to the binary star.}
{The resulting masses for components Aa of $14.0\pm2.2$ $M_\odot$ and Ab
of $7.4\pm1.1$ $M_\odot$ are low compared to theoretical expectations,
with a distance of $294\pm21$ pc which is smaller than a photometric
distance estimate of $387\pm54$ pc based on the spectral type B0III
of the B component.  If the latter (because it is also consistent
with the distance to the Orion OB1 association) is adopted, the mass
of the secondary component Ab of $14\pm3$ $M_\odot$ would agree with
classifying a star of type B0.5IV. It is fainter than the primary by
about $2.2\pm0.1$ magnitudes in the visual. The primary mass is then
determined to be $33\pm10$ $M_\odot$.  The possible reasons for the
distance discrepancy are most likely related to physical effects, such
as small systematic errors in the radial velocities due to stellar winds.}
{}


\titlerunning{Dynamical mass of Zeta Orionis Aa}

\authorrunning{C.~Hummel et al.}

\maketitle

\section{Introduction} \label{introduction}

Studies of double-lined eclipsing binaries have been very
successful in measuring masses of stars on the main sequence
\citep{2010A&ARv..18...67T} with high enough accuracies to challenge
stellar evolution models. The latter have held up well even for
O-type stars \citep{2003IAUS..212...91G}.  However, even though
O-type supergiants have also been found in eclipsing binaries, the
observational selection bias that favors closer systems over wider
ones, would indicate a high probability that the components in these
systems have interacted and therefore would not be described by 
single-star evolution models.

High-angular resolution techniques based on optical long baseline
interferometers have overcome this limitation and have contributed
significantly to the stock of precise stellar mass measurements.  Here we
report on our attempt to make the first such determination for a supergiant 
in a detached system, excluding any mass transfer.

\citet{2000ApJ...540L..91H} found a companion 40 mas from the O9.5\,Ib
component A (HR 1948) of the wide double $\zeta$ Orionis AB, and
detected orbital motion over the course of a few months. Using the Navy
Precision Optical Interferometer\citep{1998ApJ...496..550A}, we continued
observations over the 7.3 year orbital period in order to determine the
orbital elements.  Component B (HR 1949) showed slow motion relative to
A by increasing the position angle 9.6 degrees over the past century,
and by decreasing the separation by 116 mas at the same time. It is
currently 2.40 arcsec away from the primary at a position angle of 165
degrees (Washington Double Star Catalog).  We interpret this as orbital
motion in a common proper motion physical pair.

A detailed spectroscopic study of $\zeta$ Orionis A was carried out by
\citet{2008MNRAS.389...75B}, who determined an effective temperature
of $T_{\rm eff}=29500\pm1000$ K for component Aa (the secondary,
Ab, was not taken into account) and $\log{g}=3.25\pm0.10$ with normal
abundances. The rotational velocity (of the primary) was determined to
be $v\sin{i}=110\pm10$ km/s, and the rotation period to seven days, implying
an inclination of the rotation axis of $40^\circ$.

Meanwhile, we were able to identify photospheric lines of both the primary
and secondary components in archival spectra obtained with HEROS and
FEROS at La Silla, Chile, with BESO at Cerro Armazones, Chile,
FOCES at Calar Alto, Spain, and new observations with UVES at the VLT.
This allowed us to determine a dynamical mass and distance to $\zeta$
Orionis A.

\section{Observations and data reduction} \label{observations}

\subsection{Interferometry}

Observations prior to 2000 were described in \citet{2000ApJ...540L..91H}.
While these and the observation in 2000 used the three-way beam combiner,
subsequent observations used the six-way beam combiner described
by \citet{1998ApJ...496..550A}. In Table~\ref{psn} we list the dates
and stations used (along with astrometric results described
later).  For a given combination of stations (``configuration'', per
spectrometer), the signals of all baselines thus realized were decoded
and Fourier-transformed to yield complex visibilities as described by
\citet{2003AJ....125.2630H}. For the observations in the years of 2004
and later, fewer than the maximum number of stations were allowed to
illuminate a spectrometer owing to crosstalk between the fringe signals
at different modulation frequencies.

The data were reduced as described by \citet{2003AJ....125.2630H}.
Additional ``incoherent'' scans away from the fringe packet were executed
for each star and each station configuration in order to precisely
determine the visibility bias. Calibration was performed using interleaved
observations of $\epsilon$ Orionis adopting a limb-darkened
diameter of 0.9 mas \citep{1991AJ....101.2207M}, for each configuration
separately. Calibration uncertainties based on scan-to-scan variations in
the calibrator visibility range from a few percent up to 20\%, and in the
closure phase up to a few degrees. Since channel-to-channel variations of
the visibility amplitude spectra are much lower in value, we allowed
the calibration of each spectrum to float during the astrometric
fits rather than applying the larger calibration uncertainties. The
(limb-darkened) diameter of $\zeta$ Orionis Aa was determined to be 0.58
mas by \citet{1982A&A...105...85R} based on the infrared flux method,
and measured to be $0.48\pm0.04$ mas by \citet{1974MNRAS.167..121H} using
an intensity interferometer. The latter group already noticed the presence
of a third component in the $\zeta$\,Ori AB system, and even predicts
a magnitude difference of about 2\,mag. However, they state that the
stellar diameter derived from intensity interferometry should be taken as
trustworthy for component Aa, since the contributions from the additional
components would be much smaller than the uncertainty. We computed an
additional estimate of the diameter based on a B0I template SED fit to
archival photometry, giving a value of $0.40\pm0.02$ mas. We therefore
adopted a ``mean'' value of 0.48 mas, but such a diameter is
mostly unresolved by our interferometric observations.

\begin{figure}
\centering
\includegraphics[width=1.0\columnwidth]{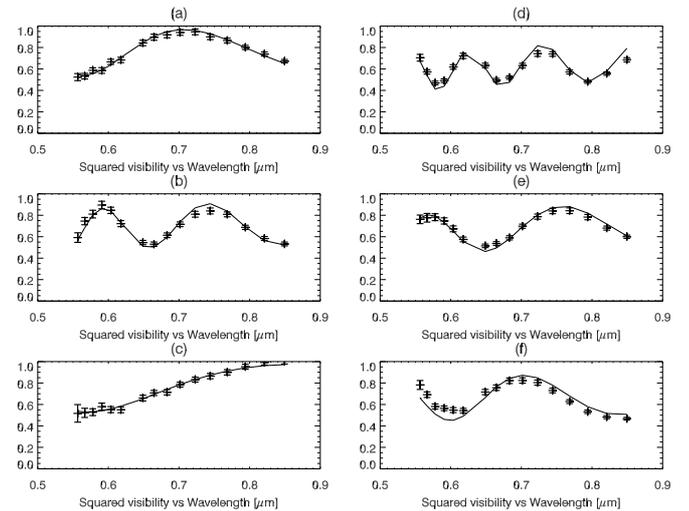}
\caption{Calibrated (squared) visibility amplitudes plotted versus
wavelength for 2002 Dec 20 on the E-E2(a), E2-W(b), E2-N(c), E-W(d),
E-N(e), and N-W(f) baselines at 7:45 UT.  The solid line shows the model
prediction for a fit with component separation $\rho=24.6$\,mas and PA
$\theta=87.7^\circ$. The amplitude of the quasi-sinusoidal amplitude
variation is fit with a magnitude difference $\Delta m=2.2$.
}\label{vissq}
\end{figure}

As an example, we show visibility spectra of all six baselines in a full
four-station observation in Fig.~\ref{vissq}. The full coverage of the
sinusoidal visibility variation demonstrates that the precision of the
astrometric fits should in fact benefit from a floating calibration.

\begin{table}
\caption{Observation and result log. Columns 1 and 2 give the date and
Julian year, col.\ 3 the stations (astrometric 
stations C, E, W, and N, as well as imaging stations E2, E6, and W7), 
columns 4 and 5 separation (mas) and position angle (deg), respectively,
and columns 6 to 8 major and minor axes and postion angle of the uncertainty
ellipse for the derived relative position of binary components.}
\label{psn}
\begin{flushleft}
\begin{tabular}{cccccccc}
\hline\hline
Feb 13& 1998.12&  CEW        & 42.1& 276.3&  0.60&  0.13&175\\
Feb 14& 1998.12&  CEW        & 42.1& 276.4&  0.60&  0.13&176\\
Mar 03& 1998.17&  CEW        & 42.5& 275.1&  0.58&  0.13&175\\
Mar 20& 1998.21&  CEW        & 43.0& 273.9&  0.58&  0.15&176\\
Nov 19& 1998.88&  EW         & 46.7& 260.0&  2.40&  0.11&163\\
Nov 26& 1998.90&  CEW        & 46.8& 258.8&  0.56&  0.16&178\\
Feb 13& 1999.12&  CW         & 47.3& 254.6&  2.20&  0.19&155\\
Feb 17& 1999.13&  CW         & 47.2& 254.2&  2.13&  0.18&155\\
Feb 18& 1999.13&  CW         & 47.2& 254.0&  1.65&  0.20&153\\
Feb 23& 1999.14&  CEW        & 47.0& 253.8&  0.58&  0.14&176\\
Mar 29& 1999.24&  CEW        & 47.1& 251.7&  0.58&  0.19&179\\
Mar 30& 1999.24&  CEW        & 47.1& 251.6&  0.55&  0.16&177\\
Oct 18& 2000.80&  CEW        & 37.3& 214.1&  0.64&  0.11&177\\
Jan 09& 2002.02& EWNE2       & 26.2& 154.7&  0.25&  0.13&146\\
      &        & CWE2W7      &     &      &      &      &   \\
Jan 12& 2002.03& EWNE2       & 26.1& 153.8&  0.25&  0.15&146\\
      &        & CWE2W7      &     &      &      &      &   \\
Dec 20& 2002.97& EWNE2       & 24.5&  87.7&  0.20&  0.15&176\\
      &        & CWE2        &     &      &      &      &   \\
Mar 11& 2004.19& CE, CW      & 23.5& 332.2&  0.49&  0.28&178\\
Mar 12& 2004.19& CE, CW      & 23.6& 332.5&  0.48&  0.29&  6\\
Feb 24& 2006.15& CEN         & 46.4& 261.5&  0.28&  0.15& 23\\
      &        & CWE6        &     &      &      &      &   \\
\hline
\end{tabular}
\end{flushleft}
\end{table}

We determined the separation, $\rho$, and position angle, $\theta$,
of the binary components for each night from the visibility data,
starting with estimates derived from images made using standard
phase self-calibration and deconvolution methods \citep[see examples
in][] {2000ApJ...540L..91H}. This procedure enabled unambiguous
identification of the global minimum of $\chi^2$ corresponding to the best
fit values of $\rho$ and $\theta$ to the visibility data. The astrometric
fit results are also listed in Table~\ref{psn}. The uncertainty
ellipses correspond to one-seventh of the synthesized beam, which has
been shown to give realistic estimates of the astrometric accuracy of
NPOI multi-baseline observations. Using the orbit determined with these
measurements (as described below), a simultaneous fit to all visibility
data determined the magnitude difference to be $2.2 \pm 0.1$ over the
NPOI band width ($0.70\pm0.15 \mu$m) without significant color changes.

\subsection{Spectroscopy}

Spectroscopic observations were taken with various echelle instruments.
In 1995, 1997, and 1998 spectra were obtained with the HEROS instrument,
attached to ESO telescopes on La Silla. The HEROS observational campaigns
were described by \citet{kaufer98}, and the data format and processing
in detail by \citet{1995JAD.....1....3S}. In brevity, HEROS provided
spectra with a spectral resolution of $R=20\,000$ from about 350 to
about 870\,nm.  For individual spectra a $S/N$ above 100 was aimed
for. For $\zeta$\,Ori, there were 57 spectra taken in 1995 (MJD\,49747
to 49804), 16 in 1997 (MJD\,50449 to 50457), and 64 in 1998 (MJD\,51143
to 51160). For the purpose of this work, deriving the orbital radial
velocity (RV) variations, spectra within one season have been averaged to
increase $S/N$ to almost 1000, and the mean time was taken as epoch.

FEROS data were obtained from the ESO archive. In 2006, one single
spectrum was taken at MJD=53738, and in 2009 5 spectra were obtained on
MJD=54953. FEROS data range from 370\,nm to 900\,nm with $R=48\,000$. See
\citet{kaufer99} for a more detailed description of the instrument. The
combined $S/N$ for each epoch is between 100 and 200.

A single spectrum was taken in 2001 with FOCES at the 2.2\,m telescope
of Calar Alto, Spain. The wavelength coverage ranges from 390\,nm to
950\,nm, with $R\approx 40000$. Data reduction was performed using the
FOCES data reduction software by N.\ Przybilla (priv.\ comm.).

Several spectra were recorded over three observing seasons in 2009 and
2010 by the BESO spectrometer (based on the FEROS design) at the Hexapod
1.5m telescope at Cerro Armazones, Chile.  The wavelength coverage ranges
from 370\,nm to 860\,nm, with $R\approx 48000$.

A few spectra taken in 1995, 1998, and 1999 were found in the public ELODIE 
archive\footnote{\tt http://atlas.obs-hp.fr/elodie/}
of the Haute Provence Observatory \citep{2004PASP..116..693M},
covering the wavelength range of 400\,nm to 680\,nm, at $R=42000$.

Finally, UVES spectra have been obtained in the 437+760 setting
\citep{2000SPIE.4008..534D} separately of components A and B, which
were easily resolved. A total of 32 spectra were taken of $\zeta$\,Ori
A within half an hour and averaged to a combined $S/N$ of well above
1000. Additionally, five/four (blue/red arm) spectra of $\zeta$\,Ori B
were taken, with a combined $S/N$ of about 500.

We checked the wavelength calibration to confirm the same position of the
interstellar Ca\,{\sc ii} H (3934) and Na\,{\sc i} (D$_1$) lines (5896)
in all spectra. The rms variation was 1.0 km\,s$^{-1}$ and 1.7 km\,
s$^{-1}$, respectively. However, these lines are shifted by several
times the rms in the FOCES spectrum, and the same is true for the Na
D line in all but the last epoch of the ELODIE spectra. The wavelength
calibration of these spectra should therefore be considered uncertain
so they were excluded from further analysis.

In the spectra all He\,{\sc i} lines, as well as He\,{\sc ii} 4686, have
a relatively {narrow core} with varying RV in one direction, while the
line {wings} are shifted in {anti-phase} with respect to the cores. This
is the signature of a double-lined (SB2) binary. For some lines (almost)
exclusive {formation in the O9.5 component} can be assumed. The best
candidates are the He\,{\sc ii} lines, typically not seen in B-type
stars (except He\,{\sc ii} 4686, which explains why RVs
we derived from single Gaussian profiles fitted to this line did not
show orbital motion). We also noticed systematic offsets between lines
nearly independent of the orbital phase, such as an offset of about 5
km/s between the He\,{\sc ii} lines at 4542 and 4200, which appeared
to be due to a slight non-Gaussian shape of the latter. Such offsets
were also noted by \citet{2012A&A...542A..95R} in the case of another
O-star binary. Therefore, we decided not to use multi-line fits and
derived RVs from He\,{\sc ii} 4542 with Gaussian fits to
the line. The width of this line was about 3 $\AA$, as expected from the
$v\sin{i}$ of 110 km/s. Theoretical spectra based on model atmospheres
\citep{1995ApJ...439..875H} show that this line is basically absent in
an early type B dwarf or subgiant ($T_{\mathrm eff}=27500$ K and $\log g
=4.0$). However, \citet{2008MNRAS.389...75B} note line shape variability
with a peak-to-peak maximal amplitude of 17 km/s (in particular He\,{\sc
i} 4922) on time scales of the rotational period of the primary related
to its magnetic field modulating the wind.

There are as well very weak and rather narrow lines that are not expected
in the O9\,Ib star. These narrow lines are RV variable in the same sense
as the cores of stronger lines, i.e.~they belong to the companion and are
indicative of an early type B star. Again, we fit Gaussian profiles to
the O\,{\sc ii} 4941 and 4943 lines to derive the RV of the
secondary (as a single parameter in the fit). The width of these lines was
typically 1 $\AA$. Theoretical spectra \citep{1995ApJ...439..875H} confirm
that these lines are deeper by a factor of five than the same lines in
an O9.7 supergiant ($T=30000$ K and $\log(g)=3.5$). Combined with the fact
that the primary lines are rotationally broadened, this choice for measuring
the secondary velocities is justified. The O\,{\sc ii} 4907
line was not usable in several spectra, and was therefore not included.
The results of the Gaussian-fit measurements are given in Table~\ref{vel}.

We estimated the uncertainty of each RV measurement of
the secondary from their spread when fit to simulated line profiles
designed to have the same noise statistics and residual fit deviations
from Gaussian profiles as the measured profile (Monte Carlo method).
For the lines of the primary, which had a much higher S/N, we estimated
the level of systematic error related to profile variations by comparing
the Gaussian fit results with RVs obtained by matching
the mirrored profile with itself at about half of the line depth. The
resulting estimate for the lower limit of the uncertainty related to
non-Gaussian profiles of 1.5 km/s applicable to all observations of this
line was then added in quadrature to the Monte Carlo uncertainty estimate
for each measurement.

\begin{table}
\caption{Radial velocity measurements. For the primary, only He\,{\sc ii}
was used, for the secondary O\,{\sc ii}. Measurements flagged with a : 
were not used in the final analysis (see text).}
\label{vel}
\begin{flushleft}
\begin{tabular}{ccllccccc}
\hline\hline
JD-2400000 & Phase & He\,{\sc ii} 4542 & O\,{\sc ii} 4943 & Instrument \\
49771.5    & 0.90  & 16.7 $\pm$ 1.8 & 50.0 $\pm$ 2.8 & HEROS\\
50454.5    & 0.15  & 31.0 $\pm$ 2.1 & 24.2 $\pm$ 2.6 & HEROS\\
51143.5    & 0.41  & 35.0 $\pm$ 1.7 & 14.9 $\pm$ 3.3 & HEROS\\
52182.7    & 0.79  & 25.8:          & 33.3:          & FOCES\\
53738.7    & 0.37  & 39.6 $\pm$ 1.9 & 12.8 $\pm$ 5.0 & FEROS\\
54501.5    & 0.66  & 29.7 $\pm$ 2.3 & 21.7 $\pm$ 2.3 & FEROS\\
54873.6    & 0.80  & 23.2 $\pm$ 1.6 &                & BESO\\
54919.5    & 0.81  & 23.9 $\pm$ 1.7 &                & BESO\\
55170.7    & 0.91  & 14.1 $\pm$ 1.6 &                & BESO\\
55235.5    & 0.93  & 20.3 $\pm$ 3.0 &                & BESO\\
55237.6    & 0.93  & 20.4 $\pm$ 3.7 &                & BESO\\
54954.5    & 0.83  & 23.8 $\pm$ 1.8 & 39.0 $\pm$ 2.3 & FEROS\\
55435.9    & 0.01  & 20.5 $\pm$ 1.8 & 47.3 $\pm$ 3.3 & UVES\\
50030.7    & 0.99  & 19.4:          & 42.7:          & ELODIE\\
51144.9    & 0.41  & 35.3:          &  8.0:          & ELODIE\\
51533.3    & 0.55  & 29.6 $\pm$ 2.9 &                & ELODIE\\
\hline
\end{tabular}
\end{flushleft}
\end{table}

\begin{figure}[h]
\centering
\includegraphics[width=1.00\columnwidth]{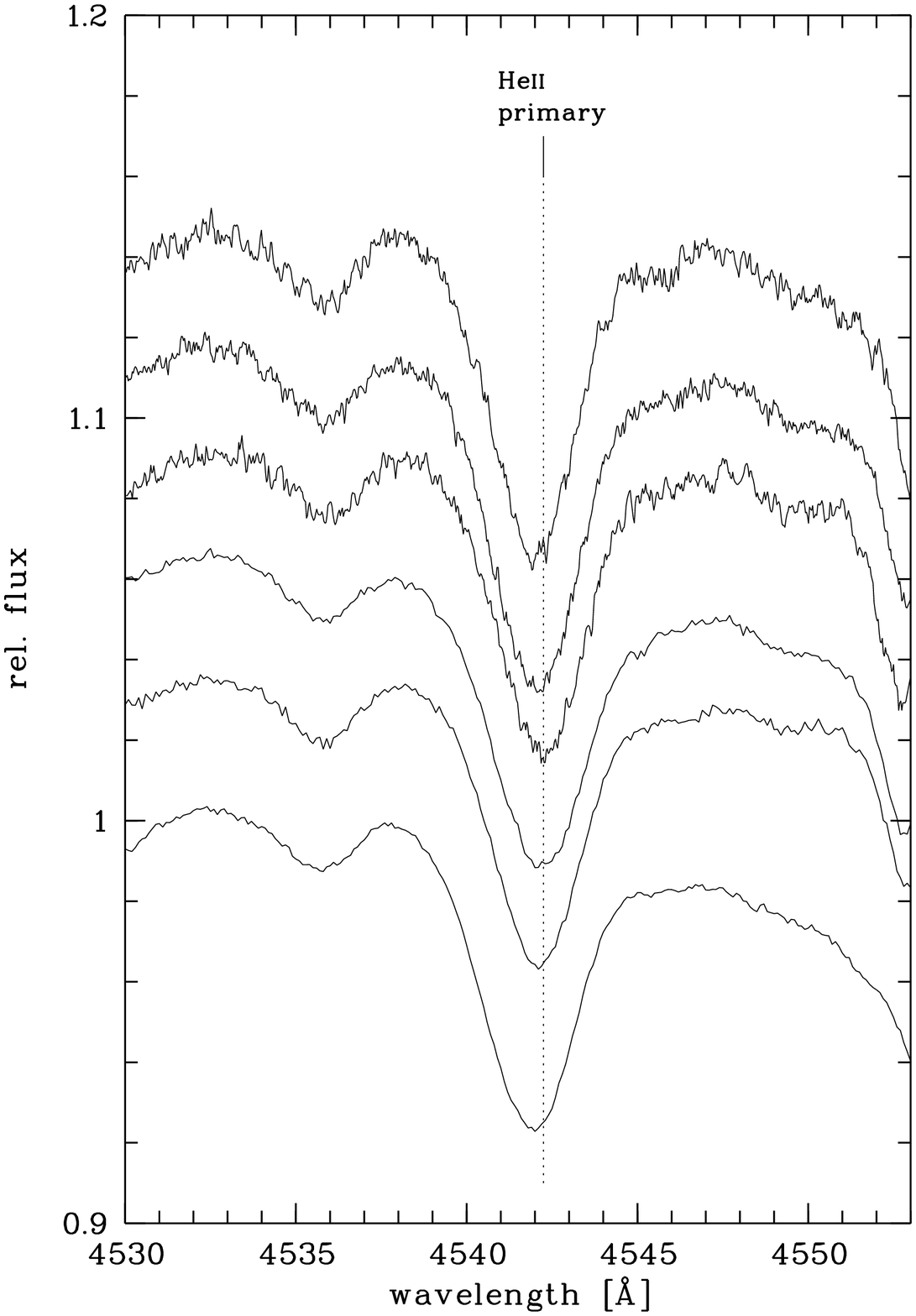}
\caption{The SB2 composite spectrum of $\zeta$ Ori Aa+Ab. The panel
shows He\,{\sc ii} 4542, presumably originating from the primary alone.
The three HEROS and three FEROS spectra are offset in increasing order
of time from the bottom up. As a guide, the RV difference between the bottom 
two spectra is 14.3 km/s.}
\end{figure}

\begin{figure}[h]
\centering
\includegraphics[width=1.00\columnwidth]{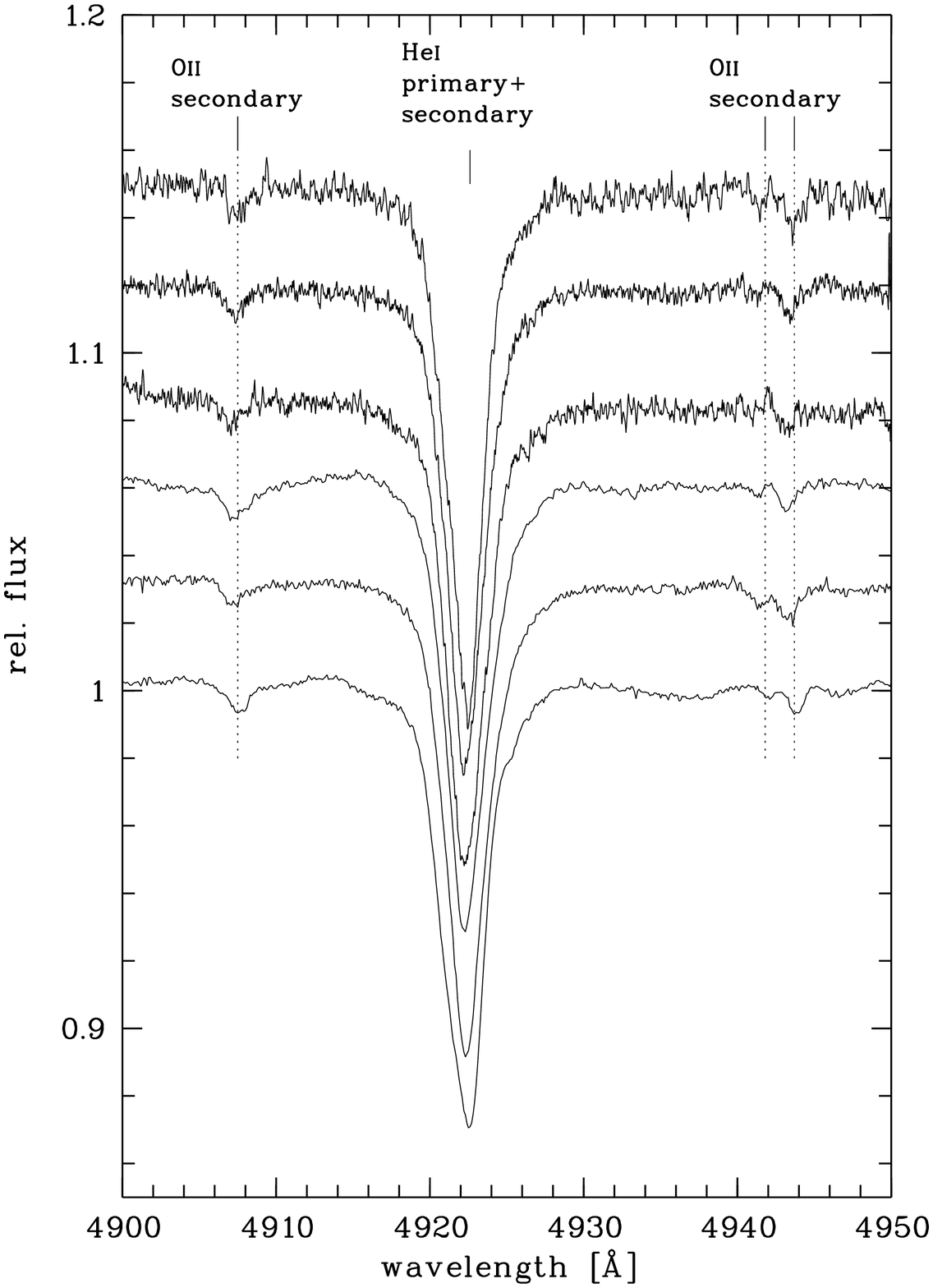}
\caption{The SB2 composite spectrum of $\zeta$ Ori Aa+Ab.  The panel
shows the weak O\,{\sc ii} 4943 lines from the secondary and the composite
He\,{\sc i} 4922 line. The three HEROS and three FEROS spectra are offset
in increasing order of time from the bottom up. As a guide, the RV 
difference between the bottom two spectra is 25.8 km/s.}
\end{figure}

\begin{table}[h]
\centering
\caption[]{Orbital elements and system parameters}
\label{elements}
\begin{tabular}{lr}
\hline\hline
Orbital period			& $2687.3\pm7.0$\,d\\
Periastron epoch 		& JD $2452734.2\pm9.0$\\
Periastron long. 		& $24.2\pm1.2^\circ$ \\
Eccentricity			& $0.338\pm0.004$\\
Ascending node			& $83.8\pm0.8^\circ$ \\
Inclination			& $139.3\pm0.6^\circ$\\
Semi-major axis			& $35.9\pm0.2$\,mas\\
Systemic velocity 		& $28.3\pm0.5$\,km/s \\
Orbital parallax 		& $3.4\pm0.2$\,mas\\
Visual magnitude difference 	& $2.2\pm0.1$\\
$M_{\rm Aa}$ 			& $14.0\pm2.2\,\rm M_\odot$\\
$M_{\rm Ab}$ 			& $7.4\pm1.1\,\rm M_\odot$\\
$K_1$ (derived)			& $10.1$\,km/s\\
$K_2$ (derived)			& $19.6$\,km/s\\
\hline
\end{tabular}
\end{table}

\section{Results and discussion} \label{results}

\subsection{Orbital elements and distance}

\begin{figure}[h]
\centering
\includegraphics[width=1.0\columnwidth]{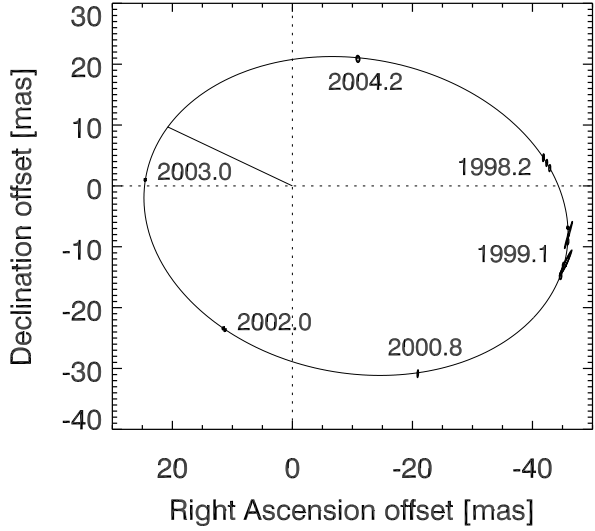}
\caption{Orbit of $\zeta$ Orionis Ab around Aa (center). The line indicates 
the secondary Ab at periastron. A few selected epochs are marked.
}\label{orbit}
\end{figure}

\begin{figure}
\centering
\includegraphics[width=1.0\columnwidth]{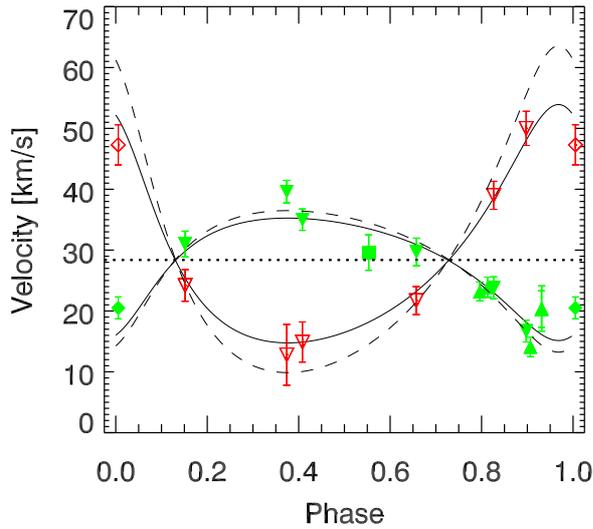}
\caption{The measured RVs of both components. The green (filled) symbols
denote the primary (He\,{\sc ii} 4542), the red (open) symbols the
secondary (O\,{\sc ii} 4943). Triangles pointing down denote FEROS/HEROS
measurements, triangles pointing up denote BESO.  Squares denote ELODIE,
and the diamond is for UVES. The dashed lines are for the model based
on the photometric distance (see discussion; the derived velocity
semi-amplitudes are $K_1=11.6$\,km/s and $K_2=26.8$\,km/s).
}\label{rvorbit}
\end{figure}

We fit the seven orbital elements, the systemic velocity, and
the stellar masses, to the astrometric positions and RVs.
Initial estimates of the elements of the apparent orbit were obtained
using the Thiele-Innes method.  Thanks to the high precision of the
astrometric orbit, the elements in common with the spectroscopic orbit did
not change much after including the RV data. The latter
mostly constrained the mass estimates.  The reduced $\chi^2$ of the fits
was about 0.5 for astrometry and 1.0 for spectroscopy. No
significant offset between the systemic velocities for primary and
secondary was found (at a level of 1 km/s). The results are summarized
in Table~\ref{elements}, and Figs.~\ref{orbit} and \ref{rvorbit} show
the fit to the measurements. An orbital parallax of $3.4\pm0.2$ mas was
derived, corresponding to a distance of $d=294\pm21$ pc and a distance
modulus of 7.4.  $\zeta$ Orionis is a member of the Orion OB1 association,
more specifically of the OB1b subgroup which is centered on the belt
stars \citep{2008hsf1.book..459B}.  The Orion star forming regions are
typically assumed to be at about 400\,pc.  \citet{2007A&A...474..515M}
derived a distance of $414\pm7$ pc from VLBA astrometry to the Orion
nebula cluster, confirmed by \citet{2009A&A...497..195K}, who obtained
an orbital parallax from one of its members.  Our measurement here of
$\zeta$\,Ori comes in low and would indicate a location on the near side
of the association.

Apart from the more general criticism of Hipparcos distances of O-stars
expressed by \citet{2004A&A...428..149S}, there are reasons particular to
the $\zeta$\,Ori system to be careful. As an astrometric binary with a
period of less than a decade, and Hipparcos measuring over three years,
a five-parameter solution, i.e.\ without taking orbital acceleration
into account, is probably not sufficient. However, in the new Hipparcos
reduction \citep{2007A&A...474..653V}, $\zeta$\,Ori data was solved with
five parameters only for a distance of $225^{+38}_{-27}$\,pc. For this
reason we discard the Hipparcos parallax for determining the distance. A
distance measurement to $\zeta$\,Ori derived from the \ion{Ca}{ii}\,H+K
equivalent width, which gives $297\pm45$\,pc \citep{2009A&A...507..833M},
is potentially problematic, because the short distance makes it
sensitive to localized density fluctuations. Despite the apparent agreement
of these results with our distance estimate, we take an alternative
approach in the following.

The distance can also be estimated photometrically, in
particular since all components contributing to the integrated light
are known now with some certainty. Str\"omgren photometry was taken
from \citet{1998A&AS..129..431H}. These data, $V=y=1.75$\,mag, are
measurements of the combined flux, i.e.\ including components Aa, Ab,
and B. \citet{1969AJ.....74..764W} lists the magnitude difference between
A and B as $\Delta V_{\rm (A-B})=2.08$\,mag, \citet{1976AJ.....81..245E}
gives $\Delta V_{\rm (A-B})=2.1$\,mag, and \citet{1969AJ.....74.1082M}
$\Delta V_{\rm (A-B})=2.2$\,mag, however referring to older sources
from the first half of the $20^{\rm th}$ century. In the following,
\citeauthor{1969AJ.....74..764W}'s value is used. The magnitude difference
between Aa and Ab of $\Delta V_{\rm (Aa-Ab})=2.2$\,mag is derived from
fitting the interferometric data.  With these values, the individual
component magnitudes are $V_{\rm Aa}=2.08$\,mag, $V_{\rm Ab}=4.28$\,mag,
and $V_{\rm B}=4.01$\,mag, while $V_{\rm Aa+Ab}=1.93$\,mag.

The colour excess is given as $E(B-V)=0.06$ by
\citet{1968ApJ...152..913L}. Although this excess was derived by assuming
a single star, since all three stars in the system are well within the
Rayleigh-Jeans tail of their SED, any correction for the presence of the
two B stars would be very small. Since \citeauthor{1968ApJ...152..913L}
notes that in the outer areas of Orion, explicitely including the
``northwestern regions'', the extinction would be normal with $R \simeq
3.0$, the photometric fluxes were dereddened with the usual $R_V=3.1$;
i.e., $A_V=0.19$\,mag.

Component B is well known to be a B0\,III star, which is also confirmed by 
our spectrum. The B2\,III classification by \citet{1976AJ.....81..245E}
comes from a photometric classification scheme utilizing the magnitude
difference, but assuming the combined magnitude of the Aa+Ab subsystem is
due to component Aa alone. Since Ab was not known then, this introduced
a bias.

Assuming an absolute magnitude for a B0\,III star of $M_V=-4.12$\,mag
\citep{2001ARep...45..711L,2013A&A...550A..26N} and the above values for
$V_{\rm B}$ and $A_V$, the distance modulus becomes 7.94\,mag. Taking
0.3\,mag as combined uncertainty for the distance modulus from $M_V$,
$m_V$, and $A_V$, the photometric distance to the $\zeta$\,Ori system
is $d=387\pm54$\,pc, i.e.\ the photometric parallax is $\pi_{\rm
phot}=2.6^{+0.4}_{-0.3}$\,mas.  This value is only marginally consistent
with the orbital parallax.

\subsection{Stellar parameters}

We note that for the photometric distance the absolute magnitudes for
components Aa and Ab would point to spectral types of O9.5\,Iab and
B0.5\,IV, in good agreement with the spectroscopic evidence. As the
photometric distance is based on a reliable spectral classification of a
non-supergiant for which recent calibrations of the absolute magnitude
exist \citet{2013A&A...550A..26N}, we regard the photometric distance
as more robust than the orbital parallax, which is based on rather low
velocity amplitudes. We therefore fit the stellar masses to the combined
interferometric and spectroscopic data under the condition that the
results are consistent with the photometric distance. The results are
given in Table~\ref{tab_stelpam}. The mass estimates are now significantly
higher than those given in Table~\ref{elements}, but in particular for
the secondary to be more in line with what is expected  for a star of its type
and luminosity class \citep{2010A&ARv..18...67T}.  As far as the mass of
the supergiant, no previous dynamical measurements of O-type supergiants
exist with periods long enough to exclude a history of mass exchange.

\begin{table}
\caption[xx]{\label{tab_stelpam}Stellar parameters for the components of the
$\zeta$\,Ori system based on the photometric distance. 
$^a$adopted (see text).}
\begin{center}
\begin{tabular}{lccc}
Parameter         & Aa & Ab & B \\
\hline\\[-2ex]
Sp.\ type  			& O9.5\,Iab  	& B1\,IV 	& B0\,III \\
$m_V$  [mag] 			& $2.1$ 	& $4.3$  	& $4.0$ \\
$M_V$ (photometry) [mag]	& $-6.0$ 	& $-3.9$  	& $-4.1^a$ \\
$M_V$ (orbit) [mag] 		& $-5.5$ 	& $-3.3$  	& $-3.6$ \\
$M_\star$ [$\rm M_\odot$]     	& $33\pm10$ 	& $14\pm3$   	& -- \\
$R_{\star}$ [$\rm R_\odot$] 	& $20.0\pm3.2$ 	& $7.3\pm1.0$	& -- \\
\end{tabular}
\end{center}
\end{table}

As shown in in Fig.~\ref{rvorbit}, it appears that the problem
lies mostly with the velocities of the secondary, which were derived from
the O\,{\sc ii} 4943 line. As mentioned before, a small contribution of the
primary to this line can be expected based on theoretical atmosphere models,
which is why it would be premature to claim that the stellar masses derived
from the RV curves alone would point to lower values then expected for the
supergiant. Since we do not have enough very high S/N spectra
(such as the UVES spectrum) to attempt spectral disentangling (e.g., 
using FDBinary by \citet{2004ASPC..318..111I}), we adopt the results 
that are consistent with the photometric distance. 

\subsection{Evolutionary state}

Given the apparent diameter adopted for the primary, we computed a physical
radius of $20\pm3 R_\odot$ with a parallax $\pi=2.6$ mas, consistent with
this component being a supergiant. The same value can be derived from the
effective temperature and luminosity (see Table~\ref{tab_stelpam}).

\begin{figure}
\centering
\resizebox{\hsize}{!}{\includegraphics[width=1.0\columnwidth]{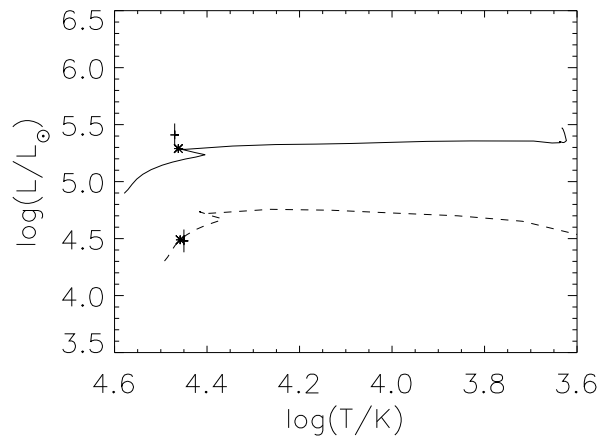}}
\caption{The location of the components ($+$ symbols) of $\zeta$
Orionis A in the theoretical plane of the HR diagram, based on
the photometric parallax of 2.6 mas and bolometric corrections
of $-2.84$ for the primary and $-2.72$ magnitudes for the secondary
\citep{1996ApJ...469..355F}. Stellar evolution tracks of a 25 $M_\odot$ 
star (solid line) and of a 15$M_\odot$ star (dashed line) are shown, and 
asterisks are used to mark the primary at an age of 6.4 Myrs and the 
secondary at an age of 7.2 Myrs.
}\label{evo}
\end{figure}

To determine the (approximate) age of $\zeta$ Orionis A, we used
the stellar evolution models computed by \citet{1992A&AS...96..269S,
1993A&AS..101..415C,1993A&AS..102..339S,1994A&AS..103...97M} for masses
of 15 and 25 times solar, for metallicities of 0.001, 0.004, 0.008,
0.020, and 0.040. For metallicities lower than solar, models matching
the derived luminosity and effective temperatures are older and place
the primary on the hydrogen-shell burning track following the exhaust
of the core hydrogen supply. However, the secondary is quite a bit older
here than the primary, and these models are therefore invalidated. Using
solar metallicity models, $Z=0.020$, we find that at an age of about
7 Myr, the modeled stars match the observed properties reasonably well,
considering that a model for the mass of the primary was not available
(Fig.~\ref{evo}). \citet{2012A&A...538A..29M} determined a younger age 
of $3.6\pm0.7$ Myr based on the combined properties of $\zeta$ Ori A 
(adopting a mass of $42M_\odot$), using models of rotating massive stars by
\citet{2011A&A...530A.115B}. 

\section{Component B} 

\begin{figure}[t]
\includegraphics[angle=270,width=8.8cm,clip]{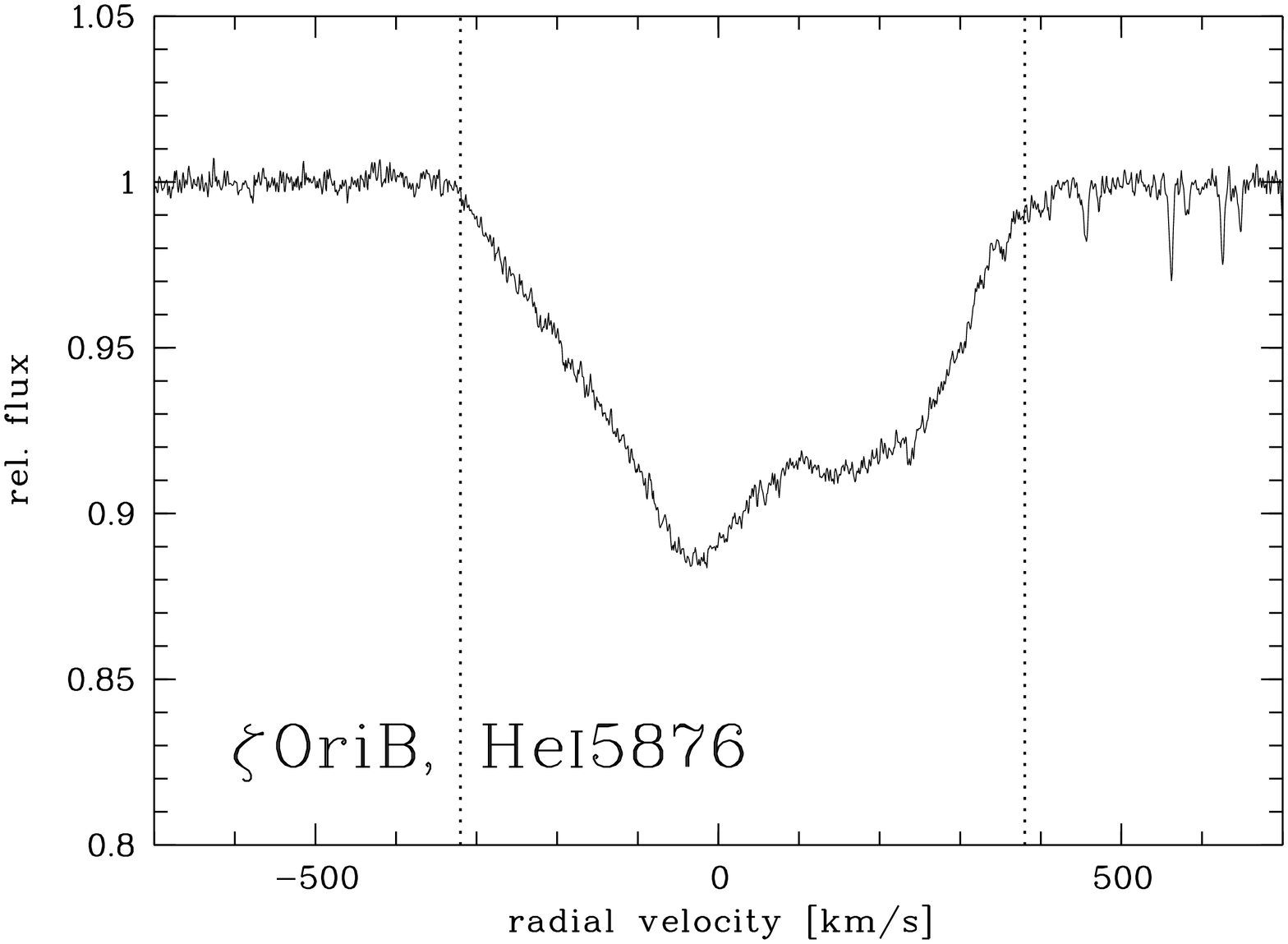}%
\caption[xx]{\label{fig_zetOriB}The \ion{He}{i}\,5876 line of $\zeta$\,Ori
B. Apart from being a very rapid rotator, the pulsational signature is
clearly seen. The dotted lines indicate $-320$ and $+380\,{\rm
km\,s^{-1}}$}
\end{figure}

To our knowledge, this is the first time that a high-quality spectrum
for component B of the $\zeta$\,Ori system has been obtained. The spectrum
is that of a very rapidly rotating star, at about $v\sin i = 350\,{\rm
km\,s^{-1}}$ and a systemic velocity of $v_{\rm sys}=25\,{\rm km\,s^{-1}}$,
the latter consistent with the $\gamma$ velocity of the Aa+Ab
subsystem. The line profiles in the spectrum are highly distorted from
a purely rotationally broadened profile (Fig.~\ref{fig_zetOriB}). Such
distortions in early type stars could either be due to abundance
patterns on the stellar surface or due to nonradial pulsation. Abundance
patterns are unlikely, since the distortion has a similar shape in
all lines, regardless of species. To produce such a shape,
nonradial pulsation has to be a high-order $p$-mode type, meaning
$\zeta$\,Ori B is a $\beta$\,Cephei star without a radial mode, where
the photometric variations mostly cancel across the surface. According
to Fig.~3 of \citet{2006A&A...452..945T}, such a line profile variation
can be found in roughly half of the early B type stars with $v\sin i
> 250\,{\rm km\,s^{-1}}$.  Other examples for such rapidly rotating
$\beta$\,Cephei stars include $\pi$\,Aqr \citep{2005ASPC..337..294P}
and $\delta$\,Sco \citep{1986ApJ...304..728S}.

\section{Conclusions}

We have measured the apparent orbit of the companion of $\zeta$ Orionis
Aa discovered by \citet{2000ApJ...540L..91H} over its seven-year period,
and determined the orbital elements. To determine a dynamical mass
of the components, we analyzed archival spectra to measure the
RV.  We determined a dynamical mass estimate (14 M$_\odot$)
for the O-type supergiant component Aa, for which any history of mass
exchange was excluded owing to the distance of the companion (minimum
of 9.5 AU). The orbital parallax, however, indicates a distance (294
pc) of the system from Earth that is shorter than our photometric
distance estimate (387 pc) based on the spectral type (B0III) of one
of the components.  Adopting the photometric distance, the mass of the
supergiant would increase to 33 M$_\odot$, closer to expectations based
on stellar models. The RVs were based on the shifts in
the \ion{He}{ii} (4542) line of the primary and the \ion{O}{ii} (4943)
line of the secondary. Considering the low-velocity semi-amplitude and
the paucity of suitable photospheric lines that can be ascribed to just
one of the components (our data do not allow disentangling), we conclude
that physical processes affecting line properties such as stellar winds
may have led to small systematic RV errors.

\begin{acknowledgements}
We thank Dr.\ Norbert Przybilla for providing the FOCES spectrum.
The Navy Precision Optical Interferometer is a joint project of the
Naval Research Laboratory and the US Naval Observatory, in cooperation
with Lowell Observatory, and is funded by the Office of Naval Research
and the Oceanographer of the Navy. The authors would like to thank Jim
Benson and the NPOI observational support staff whose efforts made this
project possible.  This research made use of the NASA ADS abstract
service and the SIMBAD database, operated at the CDS, Strasbourg, France.
This research also made use of the Washington Double Star Catalog 
maintained at the U.S.\ Naval Observatory.
\end{acknowledgements}

\bibliographystyle{aa}
\bibliography{references,zetOria,zetOrib}

\end{document}